\documentclass[12pt,preprint2]{aastex}
\usepackage{emulateapj5,apjfonts,onecolfloat5}

\slugcomment{Received 2002 May 1; Accepted 2002 June 17; Published 2002 June 26} 

\shorttitle{Cyclotron Resonance Features in SGR~1806--20}
\shortauthors{Ibrahim et al.}

\begin{document}
\twocolumn[

\title{Discovery of Cyclotron Resonance Features in the Soft Gamma Repeater SGR~1806--20}
\author{Alaa I. Ibrahim, \altaffilmark{1,2} 
Samar Safi-Harb,\altaffilmark{3,4} 
Jean H. Swank,\altaffilmark{1} 
William Parke,\altaffilmark{2} 
Silvia Zane,\altaffilmark{5} 
Roberto Turolla\altaffilmark{6}}
\affil{
$^{1}$NASA Goddard Space Flight Center,
Laboratory for High Energy Astrophysics, Greenbelt, MD 20771, USA; 
Alaa@milkyway.gsfc.nasa.gov \\
$^{2}$Department of Physics, The George Washington University, Washington, D.C., USA \\
$^{3}$Department of Physics \& Astronomy, U. of Manitoba, Winnipeg MB, Canada R3T 2N2 \\
$^{4}$NSERC Fellow; also Universities Space Research Association \\
$^{5}$Mullard Space Science Laboratory, University College London, UK \\
$^{6}$Department of Physics, University of Padova, Italy \\
\vspace{.2cm}
{\em Published in the 2002 July 20 issue of the Astrophysical Journal Letters, 574, L51}}

\begin{abstract}

We report evidence of cyclotron resonance features from the 
Soft Gamma Repeater SGR~1806--20 in outburst, detected with the
{\em Rossi X-ray Timing Explorer} in the spectrum of a long, complex 
precursor that preceded a strong burst. The features consist of a narrow 
5.0 keV absorption line with modulation near its second and third harmonics (at 11.2 
keV and 17.5 keV respectively).
The line features are transient and are detected in the harder part 
of the precursor. The 5.0 keV feature is strong, with an equivalent width of $\sim500$ eV 
and a narrow width of less than 0.4 keV.
Interpreting the features as electron cyclotron lines in the context of 
accretion models leads to a large mass-radius ratio ($M/R > 0.3\, 
M_\odot$/km) that is inconsistent with neutron stars or that requires a low 
$(5-7) \times 10^{11}$ G 
magnetic field that is unlikely for SGRs. The line widths are also narrow 
compared with those of electron cyclotron resonances observed so far in X-ray pulsars. 
In the magnetar picture, the features are plausibly explained as ion 
cyclotron resonances in an ultra-strong magnetic field that have recently 
been predicted from magnetar candidates. 
In this view, the 5.0~keV feature is consistent with a proton 
cyclotron fundamental whose energy and width are close to model predictions. 
The line energy would correspond to a surface magnetic field of $1.0 \times 10^{15}$~G 
for SGR~1806--20, in good agreement with that inferred from the spin-down measure in 
the source.

\end{abstract}


]

\section{Introduction}

Soft Gamma Repeaters (SGRs) are a unique class of slowly rotating pulsars 
($P \sim 5-8$ s)
that glow quietly in X-rays (luminosity $L \sim 10^{35}-10^{36}$~erg$\,{\rm s}^{-1}$)
for several years and suddenly become vigorously active for
a few weeks to months, emitting hundreds of short ($\sim 0.1$~s),
bright ($L \sim 10^{39}-10^{42}$~erg$\,{\rm s}^{-1}$) bursts of soft
$\gamma$-rays (see e.g. \citealt{hu:2000} for reviews).
Occasionally, SGRs also emit giant bursts that last for up to a few 
hundred seconds and exhibit remarkable pulsations that reveal 
their spin periods and confirm their nature as rotating neutron stars 
(\citealt{ma:1979}; \citealt{hua:1999}; \citealt{ib:2001}).

The lack of evidence for a binary companion or a
remnant accretion disk made the energy source of SGRs elusive.
Two competing models offer contrasted views on SGRs as conventionally
magnetized ($B \approx 10^{12}$~G) neutron stars powered by
fossil accretion disks (\citealt{mar:2001}), or ultra-magnetized ($B \approx 10^{15}$~G) 
neutron stars powered by their own magnetic field 
({\em magnetars}; \citealt{du:1992}). Recently, strange/quark star models were also 
proposed for SGRs (\citealt{cd:1998}; \citealt{bz:2000}).

SGR~1806--20 is one of the four confirmed SGRs; three lie within
the Milky Way (SGR~1900+14, SGR~1806--20, and SGR~1627--41) and one in
the Large Magellanic Cloud (SGR 0526--66). The source was first detected by the
{\em Prognoz 7}, {\em ISEE} and {\em KONUS} instruments in 1979 \citep{la:1986}.
A few years later, it underwent several periods of burst activity that 
were observed by different missions; this allowed a fairly accurate 
determination of the source position (\citealt{att:1987}; \citealt{la:1987}).
The latest episode of activity was closely monitored in 1996 with the
{\em Rossi X-ray Timing Explorer (RXTE)}. During these observations
the source in quiescence was found to pulsate with a 7.47~s 
spin period and to spin-down at a high rate (2.6~ms$\,{\rm yr}^{-1}$).
Interpreting the spin-down as being due to magneto-rotational dipole
losses leads to a magnetic field of $\sim 8\times 10^{14}$~G and a 
characteristic age of $\sim 1500$~yr, typical of a young pulsar
(\citealt{ko:1998}). SGR~1806--20 has been associated
with the Galactic radio supernova remnant (SNR)~G10.0--0.3 about 14.5~kpc 
away (\citealt{ku:1994}; \citealt{co:1997}); however, 
this association, like those of other SGRs with SNRs (see \citealt{ga:2001}),
was later questioned on the basis of {\em IPN} and {\em Chandra} observations                          
\citep{hub:1999}. Recently, a possible infrared counterpart 
was reported on the basis of {\em Chandra} observations \citep{eik:2001}.
  
To date, while a great deal of evidence has gathered in favor of the 
magnetar model over other scenarios, the determination of the magnetic field 
strength in SGRs (\citealt{ko:1998}; \citealt{ko:1999}), although compelling, 
is still indirect (\citealt{mar:1999}). Spectral signatures, on the other hand, 
promise decisive direct measurements of the field strength. In this {\em Letter}, we report 
evidence of absorption features in the X-ray spectrum 
of SGR 1806--20. The nearly harmonic line spacing is suggestive of a 
cyclotron origin. We use the line energy to derive the star's magnetic 
field and discuss the implications of this finding on current SGR models.

\clearpage

\section{Observation and Data Analysis}

SGR 1806--20 entered an intense phase of bursting
activity in November 1996 during which it was extensively monitored
with {\em RXTE}. Proportional Counter Array (PCA) data were extracted
from the HEASARC archives and the
high-resolution event mode data were used to construct the burst light curves.
Among the bursts detected, we found one event that shows an unusual temporal
profile with a long ($\approx$~0.5 s) multi-peak precursor followed by a 
bright burst (see Figs. 1a and 1b). The time history of this event shares a number of
similarities with a remarkable burst from SGR 1900+14 where a 6.4~keV
emission line was recently discovered in the burst precursor
(\citealt{st:2000}; \citealt{ib:2001}).
In fact: (1) both bursts are preceded by a complex multi-peaked
precursor, and both precursors have similar temporal profiles that are
different from typical SGR bursts;
(2) both precursors are significantly longer than typical ($\sim 0.1$~s) 
SGR bursts; and (3) spectral evolution is detected in both precursors, while most bursts 
show a uniform spectrum.

Motivated by these similarities, we investigated in detail the spectral
characteristics of the precursor. In order to subtract the background, we
used a pre-burst persistent emission segment that is free from any
bursting activity.
We divided the precursor into four intervals and used the PCA data below 30 keV 
to fit the spectra of the different peaks separately. An absorbed power-law
model provided a qualitatively acceptable fit to the data and
revealed a moderate spectral evolution as shown in Fig.~1c. The spectrum is
harder in the second interval with a larger $\chi^2$ than in the other
three intervals. Absorbed bremsstrahlung models gave similar fits, with $kT\sim30$ keV.
The fit residuals (Fig. 2) show no significant structure 
in intervals 1, 3, and 4, but in interval 2, they suggest the presence of a narrow
absorption feature near 5~keV. The fit was improved by adding a cyclotron
absorption component, and the F-test gives a probability of $1.3 \times 10^{-3}$
for a chance reduction of the F-statistic by 6.8 for a 5.0~keV
absorption feature with a narrow width of 0.24~keV. The feature is significant
at $3.2~\sigma$.

Further inspection of the spectrum and the fit residuals suggested additional modulation 
around the second and third harmonics of the 5.0~keV feature and near 7~keV. 
The fit continued to improve with the successive addition of three absorption lines.
The energies derived from the fit are 7.5, 11.2, and 17.5~keV
respectively. With all four lines included, the F-test gives an improved
random probability of $2.2 \times 10^{-4}$ for a chance reduction of the 
F-statistic by 6.2, corresponding to a $3.7~\sigma$ significance.

The spectrum and the predicted counts of the best-fit model are shown in
Fig.~3 (bottom) along with the incident photon spectrum that would be implied by
the model (top). The fit parameters for the four intervals and the
characteristics of the line fits are given in Tables 1 and 2.
The F-statistics given there are all in
comparison to the absorbed power-law fit with no cyclotron resonance features.

Absorption features did not improve the fits significantly for the other
three intervals. We studied the implications of instrumental
effects in the PCA elsewhere and we showed that for count rates less than $9
\times 10^4$~counts/s, pile-up and dead-time effects are not
sufficient to modify the spectrum nor can they produce a spectral 
feature (\citealt{ib:2001}; \citealt{st:2000}). 
The count rate during the main burst peaks exceeded this limit, and we
therefore excluded them despite some evidence for absorption around 5 keV.
The presence of the features in an interval with a moderate count rate
and the fact that the features are present in part of the precursor 
and are not seen in intervals with comparable count rates and similar
number of counts strongly argue against significant instrumental effects. 
Besides, when fitted with a Gaussian, the 5.0 keV feature is strong, with an equivalent 
width of $\sim500$ eV, even larger than that of the 6.4 keV emission line from SGR~1900+14. 
This cannot be interpreted as being due to imperfections in the PCA response matrix. 
Numerical simulations also showed that random fluctuations could
produce power-law spectra with a $\chi^2$ as high as that of interval 2 in
about 16\% of the cases, but the fit residuals are randomly distributed
and never showed a systematic structure similar to that of the real data.

%----------------  Burst  -------------------
\begin{center}
\includegraphics[scale=0.5]{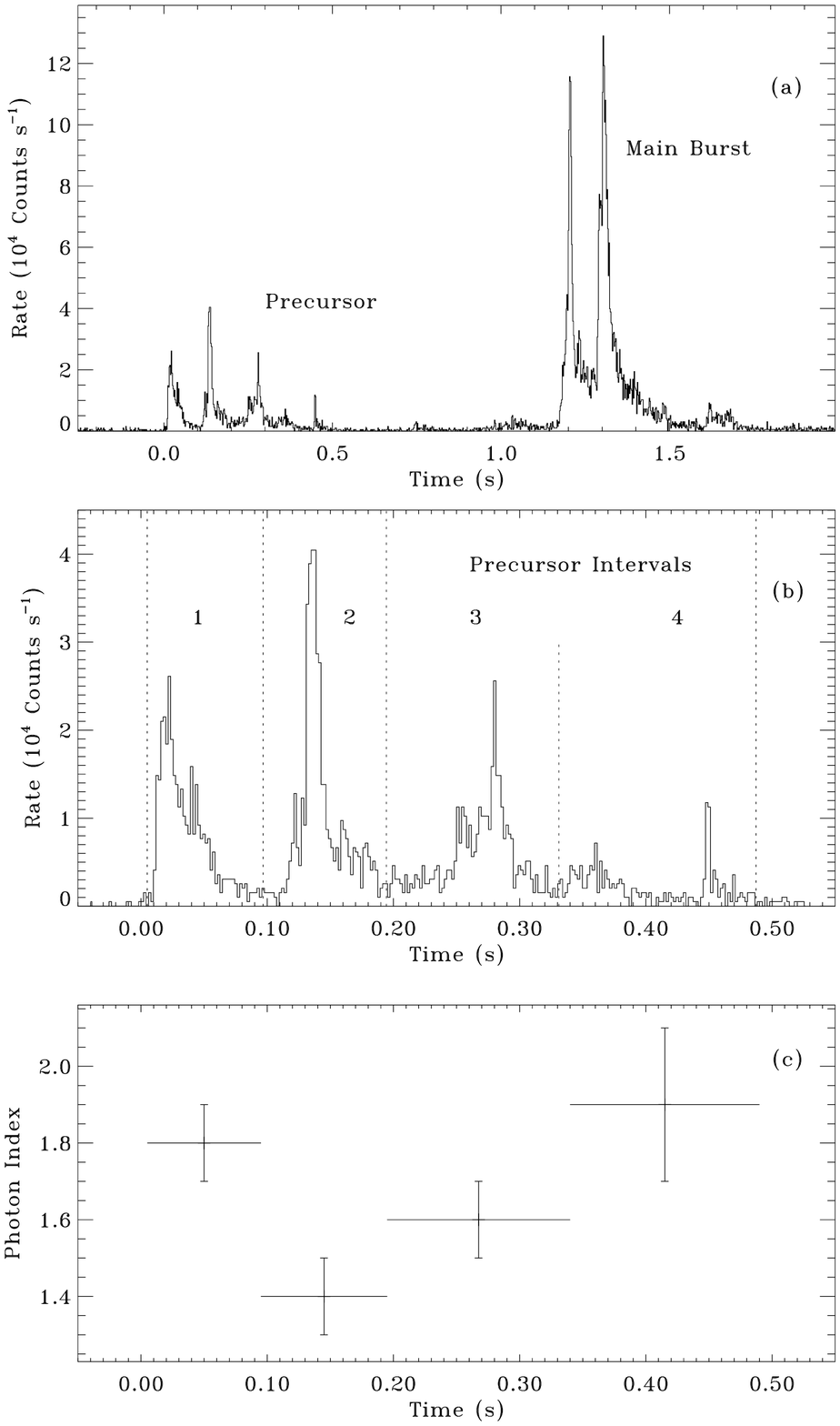}
\figurenum{1}
\figcaption{Time history of the outburst from SGR~1806--20 as seen by {\em RXTE}/PCA 
(2-60 keV). The top panel (a) shows a bright burst preceded by a 
long, complex precursor. Panel (b) shows the precursor 
intervals used in the spectral analysis. Panel (c) shows the evolution of 
the power-law photon index for the precursor intervals. $T_0(UTC)=$ 9:13:43 on 1996 Nov. 18.}
\end{center}

\begin{deluxetable}{c c c c c}
\tablecaption{Spectral Results For the Precursor}
\tabletypesize{\scriptsize}
\tablewidth{0pt}
%\tablecolumns{5}
\tablehead{
\colhead{Interval}  & \colhead{Photon Index} & 
\colhead{$N_H (10^{22} \, {\rm cm}^{-2})$}
  & \colhead{$\chi^2$/dof} & \colhead{F-statistic/Confidence Level (\%)}}
\startdata
1 & $1.8 \pm 0.2$ & $21.4 \pm  4.1$ &  35/32 & \\
2 & $ 1.4 \pm 0.1$ & $18.8 \pm 3.7$ & 40/32 & \\
3 & $ 1.6 \pm 0.1$& $18.2 \pm 3.5 $ &  12/32 & \\
4 & $ 1.9 \pm 0.2$ & $26.7 \pm 7.5$ & 31/32 & \\
2 (1 line) & $ 1.5 \pm 0.2$ & $18.7 \pm 3.4$& 23.5/29 & 6.8 / 99.87\% \\
2 (4 lines)$^{(a)}$ & $ 1.6 \pm 0.2 $ &
$21.1 \pm 4.9$ & 13/24 &  6.2 / 99.97\% \\
\enddata
\tablecomments{$^{(a)}$: The line widths are held fixed at the best-fit values shown in table 2.}
\end{deluxetable}

\begin{deluxetable}{cccc}
\tabletypesize{\scriptsize}
\tablewidth{0pt}
\tablecaption{Best-Fit Parameters for the Line Features}
\tablehead{
\colhead{Line Feature} & \colhead{Energy (keV)} & \colhead{Width (keV)} & 
\colhead{Depth}}
\startdata
1 & $ 5.0 \pm 0.2 $ & $ 0.24\pm 0.1 $ & $ 1.9 \pm 0.6$ \\
2 & $ 7.5  \pm 0.3 $ & $ 0.45 \pm 0.2 $ & $ 1.2 \pm 0.4 $ \\
3 & $ 11.2 \pm 0.4 $ & $ 1.2 \pm 0.5 $ & $ 0.9 \pm 0.3 $ \\
4 & $ 17.5 \pm 0.5  $ & $ 1.1 \pm 0.7 $ & $ 1.0 \pm 0.4 $ \\
\enddata
\tablecomments{Cyclotron absorption model is described in Mihara, T. et al. 1990, \nat, 346, 
250}
\end{deluxetable}

Taking into account that the features appear in one spectrum out of                        
the four examined, the chance probabilities for the single feature and 
the set of features are $5.2 \times 10^{-3}$ and $8.8 \times 10^{-4}$, 
respectively. However, we have now looked at a large sample of bursts 
and found the 5~keV feature in the spectra of some other bursts as well.
This will be reported in a follow-up paper (\citealt{ib:2002}). 

Since an emission line has been detected in SGR 1900+14 in a similar
event, we checked this possibility also for present data. Visual inspection
of the spectrum may be actually suggestive of a peak around 7 keV. However,
the addition of a Gaussian emission line provided a marginal improvement 
in the fit with respect to an absorbed power-law, with a confidence 
level of only 74\%.

%-------------  Residuals  --------------------
\begin{center}
\includegraphics[scale=0.24]{f2a.ps}
\includegraphics[scale=0.24]{f2b.ps}
\includegraphics[scale=0.24]{f2c.ps}
\includegraphics[scale=0.24]{f2d.ps}
\figurenum{2} 
\figcaption{Residuals of the four precursor intervals when fitted with an 
absorbed power-law model. The residuals are significant in interval 2 but are random in 
intervals 1, 3 and 4.}
\end{center}

\section{Discussion}

The presence of electron cyclotron features in X-ray
pulsars was first predicted by \citet{gn:1974} and discovered a few years
later by \citet{tr:1978} and \citet{whe:1979}. All
observed cyclotron lines have been detected above 10~keV and are
interpreted as electron features, with inferred magnetic fields 
$B \sim (1-3) \times 10^{12}$~G (\citealt{cu:1998}; \citealt{hei:1999}; 
\citealt{sa:1999}).

%-----------------  Spec  ----------------------------------------
\begin{center}
\includegraphics[scale=0.44]{f3a.ps}
\includegraphics[scale=0.44]{f3b.ps}
\figurenum{3}
\figcaption{Spectrum and best-fit continuum model for the second precursor 
interval, with four absorption lines ({\em RXTE}/PCA 2-30 keV).
{\em Bottom:} Pulse-height spectrum with the model predicted counts (histogram). 
{\em Top:} Model (histogram) and the estimated photon spectrum for the best fit model.}
\end{center}
%------------------------------------------------------------------

The presence of cyclotron features in SGRs would resolve a host of issues
concerning the nature of these peculiar objects. SGRs have been considered
as either ultra-magnetized neutron stars powered by the dissipation of their
own $B$-field (\citealt{du:1992}) or conventional neutron stars powered
by accretion from a very faint companion or a fossil disk (\citealt{mar:2001}).
Here we discuss the implications of our observation in light of these 
models.

In a strong magnetic field and depending on the physical conditions 
in the emitting/absorbing matter, cyclotron features from electrons, 
protons and other ions could in principle be observed. Possibly atomic absorption lines could 
also be present; for example, red-shifted iron at 5 keV, but no model of generating 
such high flux and high temperature at a layer deeper than absorbing heavy atoms 
has been proposed.

In a typical neutron star with $B \approx 10^{12}$~G, as invoked in 
accretion-powered models, the fundamental electron cyclotron resonance lies at 
$E_e=11.6\,(1+z)^{-1}(B/10^{12}\, {\rm G})$~keV, where 
$(1+z)^{-1} = (1-2GM/Rc^2)^{1/2}$ is
the gravitational red-shift factor at the star surface. Proton and $\alpha$-particle 
resonances, on the other hand, would be undetectable in X-rays
due to their very low energy ($E_{p,\alpha} \sim 2 - 5$~eV).  
Interpreting the 5.0~keV feature as an electron cyclotron resonance 
originating close to the surface of a typical $B \approx 10^{12}-10^{13}$~G 
neutron star would require a very large gravitational red-shift 
$(1+z )^{-1}< 0.4$ (or $z > 1.5$), with mass-radius ratio
$M/R > 0.3\, M_\odot$/km.
Such values are inconsistent with neutron stars, for which 
$M/R < 0.23\, M_\odot$/km 
as imposed by causality (\citealt{latt:1990}). Interestingly, these values 
are consistent with a more compact form of matter such as strange quark stars; 
however, no predictions for spectral 
features from such objects have been made yet (\citealt{cd:1998}; \citealt{bz:2000}).
Fields below $4 \times 10^{11}$~G are ruled out since they imply $z < 0$. 

For the 5.0~keV feature to be an electron cyclotron resonance from a neutron 
star with acceptable mass and radius, the surface magnetic field could only be in the 
narrow range $(5-7) \times 10^{11}$ G, in apparent contrast with plausible values 
for SGR~1806--20 ($B\sim10^{13}$ G) within the 
accretion model (see Fig. 2 in \citealt{ro:2000}) and well below the average value 
in ordinary radio pulsars. The pulsar $B$-field distribution shows that 
$\sim 20\%$ of the population has $B\lesssim 6\times 10^{11}$~G
(e.g. \citealt{hart:1997}; \citealt{rfp:2001}). 
The rarity of SGRs as opposed to the larger population of pulsars with such a 
magnetic field also argues against this possibility.

Moreover, electron cyclotron features in the spectra of accreting pulsars 
show large thermal broadening ($\Delta E\approx$ a few keV) because of the low
electron mass in comparison with those of ions.
Although the physical conditions in SGRs are not necessarily the same as
in X-ray pulsars, broadening of the cyclotron line is to be
expected in any case for typical electron energies $\approx 1-10$ keV.
For the 5.0 keV fundamental the estimated width would be $\Delta 
E\approx 1.5$ keV. On the contrary, the observed feature is much narrower
($\Delta E\approx 0.24$~keV). The 1~keV resolution of the PCA detector 
at 5~keV makes it difficult to
precisely resolve features with widths $\lesssim 1$~keV. However, the 90\% 
upper limit on the 5.0~keV feature width is 0.4~keV, 
still narrower than any known electron feature.
All of the preceding considerations indicate that the 
interpretation of the observed 5.0~keV feature in terms of an 
electron cyclotron line seems unlikely,
although it can not be completely ruled out on the basis of present data.

In the magnetar model ($B \gtrsim 10^{14}$~G), electron cyclotron lines lie at
very high energy $E_e \gtrsim 1$~MeV, out of range of the
{\em RXTE}/PCA, whereas proton and $\alpha$-particle resonances come within reach,
with fundamentals at $E_p = 6.3\,(1+z)^{-1} (B/10^{15}\, {\rm G})$~keV
and $E_{\alpha} = 3.2\,(1+z)^{-1}(B/10^{15} \, {\rm G})$~keV. In this picture,
the 5.0~keV feature is
plausibly a proton cyclotron fundamental with an implied magnetic field 
$B = 7.9 \times 10^{14}(1+z)$~G.  For a typical neutron star with 
$M = 1.4\,M_\odot$ and $R = 10$~km, the surface field strength in SGR 1806--20
is $B = 1.0 \times 10^{15}$~G, in a very good agreement with the value 
($B \sim 8 \times 10^{14}$~G) derived from
the spin-down measure (\citealt{ko:1998}). The features at 11.2~keV and
17.5~keV are consistent with the
second and third harmonics but with a slight deviation that may be due
to emission from a region with
different magnetic field and/or red-shift. Electron cyclotron features with
slightly anharmonic spacing have been seen in transient X-ray
sources (\citealt{hei:1999}). Another possibility is that the 17.5~keV
line is associated with the proton spin-flip transition at 
$E^S_p = 17.6\, (1+z)^{-1}(B/10^{15}\, {\rm G})$~keV 
(\citealt{tho:2000}; \citealt{za:2001}).
However, this interpretation seems less plausible, since the rates
for spin-flip transitions should contain an additional factor
$E_p/m_pc^2\ll 1$ with respect to transitions with conserved spin
projection.

In recent years the growing evidence for the existence of magnetars has
stimulated several studies on the presence of
ion cyclotron features in the X-ray spectra of SGRs
(\citealt{bez:1996}; \citealt{tho:2000}; \citealt{za:2001}; \citealt{ho:2001}).
Our results on the energy and width of the fundamental feature are
consistent with estimates by \citet{za:2001}. However, the continuum
underlying cyclotron features reported here is either non-thermal 
or bremsstrahlung for $kT>$ 20 keV and probably
originates in the star's magnetosphere. In this respect, ion
cyclotron lines may naturally form in the magnetar model of \citet{tho:1995} 
as primary photons produced in the pair fireball cross an optically 
thick baryon loaded sheath, located at $\lesssim 2 R$. The presence of 
hard photons with energy greater than the fundamental 
harmonic would also enhance the transitions between Landau 
levels and would thus makes cyclotron features stronger (\citealt{gn:1974}).

Similarly to proton cyclotron resonances, $\alpha$-particle            
resonances could also be present. For the same value of $B$ as implied by the 
observed proton fundamental, the $\alpha$-particle second harmonic would overlap the 
proton fundamental and the third harmonic would be consistent with the observed 
feature at 7.5 keV. The $\alpha$-particle fundamental would lie at 2.4~keV, too 
close to the lower end of the {\em RXTE}/PCA energy range to be separated from 
low-energy cut-off effects. Further observations with instruments possessing 
higher spectral resolution and lower bandpass energy could resolve the $\alpha$-particle 
fundamental and confirm or disprove this conjecture.

Cyclotron features offer a new diagnostic tool for probing SGRs. Further
analysis and new observations promise to shed new insights and broaden
our understanding of these enigmatic objects. This will potentially open new 
avenues for testing quantum electrodynamics predictions and exploring new 
physical effects unique to ultra-strong magnetic fields, which is one of the 
main reasons why SGRs are distinctively important.

\acknowledgments

We thank K. Hurley and T. Strohmayer for their careful reading and important comments on this 
{\em Letter}. We are also grateful to G. Pavlov, C. Thompson, R. Duncan, and C. Miller for useful 
remarks and discussions. A.I.I. thanks S.S.H and the department of Physics at Univ. Manitoba 
for their hospitality where this work was partly done. S.S.H. acknowledges support from the Natural 
Sciences and Engineering Research Council (NSERC), and by a University Research Grant (URGP) 
at the Univ. of Manitoba.


\begin{thebibliography}{}


\bibitem[\protect\citeauthoryear{Atteia et al.}{1987}]{att:1987}                           
Atteia, J.-L. et al. 1987, \apj, 320, L105                                                 
                                                                                           
\bibitem[\protect\citeauthoryear{Bezchastnov et al.}{1996}]{bez:1996}                      
Bezchastnov, V.G., Pavlov, G.G., Shibanov, Y.A., \& Zavlin, V.E. 1996,                     
in Proc. Third Huntsville Gamma-Ray Burst Symp.,  Kouveliotou C.,                          
Briggs M.F. \& Fishman J.G. Eds., AIP Conf. Proc. 384, p. 907                              
(Woodbury: New York)

\bibitem[\protect\citeauthoryear{Cheng \& Dai}{1998}]{cd:1998}                           
Cheng, K. S. \& Dai, Z. G. 1998, \prl, 80, 18
                                                                       
\bibitem[\protect\citeauthoryear{Corbel et al.}{1997}]{co:1997}                            
Corbel, S. et al. 1997, \apj, 478, 624                                                     
                                                                                           
\bibitem[\protect\citeauthoryear{Cusumano et al.}{1998}]{cu:1998}                          
Cusumano, G. et al. 1998, \aap, 338, L79                                                   
                                                                                           
\bibitem[\protect\citeauthoryear{Duncan \& Thompson}{1992}]{du:1992}                       
Duncan, R., \& Thompson, C. 1992, \apjl,  392, L9                     

\bibitem[\protect\citeauthoryear{Eikenberry et al.}{2001}]{eik:2001}                       
Eikenberry, S.S. et al. 2001, \apjl, 563, L133                                             
                                                                                           
\bibitem[\protect\citeauthoryear{Gaensler et al.}{2001}]{ga:2001}                          
Gaensler, B.M., Slane, P.O., Gotthelf, E.V. \& Vasisht, G. 2001,                           
\apj, 559, 963                                                                             
                                                                                           
\bibitem[\protect\citeauthoryear{Gnedin \& Sunyaev}{1974}]{gn:1974}                        
Gnedin, Y.N., \& Sunyaev, R.A. 1974, \aap, 36, 379                                         

\bibitem[\protect\citeauthoryear{Hartman et al.}{1997}]{hart:1997}
Hartman, J.W., Bhattacharya, D., Wijers, R. \& Verbunt, F. 1997,
\aap, 322, 477
                                                                                           
\bibitem[\protect\citeauthoryear{Heindl et al.}{1999}]{hei:1999}                           
Heindl, W.A. et al. 1999, \apjl, 521, L49                                                  
                                                                                           
\bibitem[\protect\citeauthoryear{Ho \& Lai}{2001}]{ho:2001}                                
Ho, W.C.G., \& Lai, D. 2001, \mnras, 327, 1081                                             
                                                                                           
\bibitem[\protect\citeauthoryear{Hurley}{2000}]{hu:2000}                                  
Hurley, K. 2000, in Proc. Fifth Compton Symp., McConnell M.L. \& Ryan J.M.                 
Eds., AIP Conf. Proc. 510, p. 515 (Melville: New York)                                     

\bibitem[\protect\citeauthoryear{Hurley et al.}{1999a}]{hua:1999}                          
Hurley, K. et al., 1999a, \nat, 397, 41                                                     
                                                                                           
\bibitem[\protect\citeauthoryear{Hurley et al.}{1999b}]{hub:1999}                           
Hurley, K. et al., 1999b, \apj, 523, L37                                                    
                                                                                           
\bibitem[\protect\citeauthoryear{Ibrahim et al.}{2001}]{ib:2001}                           
Ibrahim, A.I. et al. 2001, \apj, 558, 237                                                  
                                                                                           
\bibitem[\protect\citeauthoryear{Ibrahim et al.}{2002}]{ib:2002}                           
Ibrahim, A.I. et al. 2002, \apj, accepted                                                 
                                                                                           
\bibitem[\protect\citeauthoryear{Kouveliotou et al.}{1998}]{ko:1998}                       
Kouveliotou, C. et al. 1998, \nat, 393, 235                                                
                                                                                           
\bibitem[\protect\citeauthoryear{Kouveliotou et al.}{1999}]{ko:1999}                       
Kouveliotou, C. et al. 1999, \apjl, 510, L115                                              
                                                                                           
\bibitem[\protect\citeauthoryear{Kulkarni et al.}{1994}]{ku:1994}                          
Kulkarni, S.R. et al. 1994, \nat, 368, 129                                                 
                                                                                           
\bibitem[\protect\citeauthoryear{Laros et al.}{1986}]{la:1986}                             
Laros, J.G. et al. 1986, \nat, 322, 152                                                    
                                                                                           
\bibitem[\protect\citeauthoryear{Laros et al.}{1987}]{la:1987}                             
Laros, J.G. et al. 1987, \apj, 320, L111                                                   
                                                                                           
\bibitem[\protect\citeauthoryear{Lattimer et al.}{1990}]{latt:1990}
Lattimer, J.M., Prakash, M., Masak, D., Yahil, A. 1990, \apj, 355, 241

\bibitem[\protect\citeauthoryear{Marsden et al.}{1999}]{mar:1999}                          
Marsden, D., Rothschild, R.E., \& Lingenfelter, R.E. 1999, \apjl, 520, L107   

\bibitem[\protect\citeauthoryear{Marsden et al.}{2001}]{mar:2001}                          
Marsden, D., Lingenfelter, R.E., Rothschild, R.E., \& Higdon, J.C. 2001,                   
\aj, 550, 397                                                                              
                                                                                           
\bibitem[\protect\citeauthoryear{Mazets et al.}{1979}]{ma:1979}                            
Mazets, E.P. et al. 1979, \nat, 282, 587                                                   
                                                                                           
\bibitem[\protect\citeauthoryear{Regimbau \& de Freitas 
Pacheco}{2001}]{rfp:2001}
Regimbau, T. \& de Freitas Pacheco, J.A. 2001, \aap, 374, 182

\bibitem[\protect\citeauthoryear{Rothschild et al.}{2000}]{ro:2000}                        
Rothschild R.E., Marsden, D., \& Lingenfelter, R.E. 2000, in Proc. Fifth                   
Huntsville Gamma-Ray Burst Symp., Kippen R.M., Mallozzi R.S., \& Fishman                   
G.J. Eds.,  AIP Conference Series, 526, p. 842 (Melville: New York)                        
                                                                                           
\bibitem[\protect\citeauthoryear{Santangelo et al.}{1999}]{sa:1999}                        
Santangelo, A. et al. 1999, \apjl, 523, L85                                                
                                                                                           
\bibitem[\protect\citeauthoryear{Strohmayer \& Ibrahim}{2000}]{st:2000}                    
Strohmayer, T.E., \& Ibrahim, A. 2000, \apjl, 537, L111                                    
                                                                                           
\bibitem[\protect\citeauthoryear{Thompson \& Duncan}{1995}]{tho:1995}                      
Thompson, C., \& Duncan, R.C. 1995, \mnras, 275, 255                                       
                                                                                           
\bibitem[\protect\citeauthoryear{Thompson}{2000}]{tho:2000}                                
Thompson, C. 2000, in The Neutron Star-Black Hole Connection, ed.  
V. Connaughton, C. Kouveliotou, J. Van Paradijs \& J. Ventura 
(NATO-ASI), in press, astro-ph/0010016                                                          
                                                                                           
\bibitem[\protect\citeauthoryear{Tr\"umper et al.}{1978}]{tr:1978}                         
Tr\"umper, J. et al. 1978, \apjl, 219, L105                                                
                                                                                           
\bibitem[\protect\citeauthoryear{Wheaton et al.}{1979}]{whe:1979} Wheaton,                 
W.A. et al. 1979, \nat, 282, 240                                                           
                                                                                           
\bibitem[\protect\citeauthoryear{Zane et al.}{2001}]{za:2001}                              
Zane, S., Turolla, R., Stella, L., \& Treves, A. 2001, \apj, 560, 384                      
                                                                                           
\bibitem[\protect\citeauthoryear{Zhang, Xu \& Qiao}{2000}]{bz:2000}
Zhang, B., Xu, R. X., Qiao, G. J. 2000, \apj, 545, 127

\end{thebibliography}
\end{document}